\documentstyle[11pt,amssymb,epsf]{article}

\def\ben{\begin{equation}}
\def\een{\end{equation}}

\let\a=\alpha    
    
 \let\m=\mu \let\n=\nu   
     
      \let\G=\Gamma   
     
\let\C=\Chi

\def\nn{\nonumber} \def\bd{\begin{document}} \def\ed{\end{document}}
\def\ds{\documentstyle} \let\fr=\frac \let\bl=\bigl \let\br=\bigr
\let\Br=\Bigr \let\Bl=\Bigl
\let\bm=\bibitem
\let\na=\nabla
\let\pa=\partial \let\ov=\overline
\newcommand{\be}{\begin{equation}}
\newcommand{\ee}{\end{equation}}
\def\ba{\begin{array}}
\def\ea{\end{array}}
\def\ft#1#2{{\textstyle{{\scriptstyle #1}\over {\scriptstyle #2}}}}
\def\fft#1#2{{#1 \over #2}}
\def\del{\partial}
\def\vp{\varphi}
\def\sst#1{{\scriptscriptstyle #1}}
\def\oneone{\rlap 1\mkern4mu{\rm l}}
\def\td{\tilde}
\def\wtd{\widetilde}
\def\ie{\rm i.e.\ }
\def\dalemb#1#2{{\vbox{\hrule height .#2pt
        \hbox{\vrule width.#2pt height#1pt \kern#1pt
                \vrule width.#2pt}
        \hrule height.#2pt}}}
\def\square{\mathord{\dalemb{6.8}{7}\hbox{\hskip1pt}}}
\newcommand{\ho}[1]{$\, ^{#1}$}
\newcommand{\hoch}[1]{$\, ^{#1}$}
\newcommand{\bea}{\begin{eqnarray}}
\newcommand{\eea}{\end{eqnarray}}
\newcommand{\ra}{\rightarrow}
\newcommand{\lra}{\longrightarrow}
\newcommand{\Lra}{\Leftrightarrow}
\newcommand{\ap}{\alpha^\prime}
\newcommand{\bp}{\tilde \beta^\prime}
\newcommand{\tr}{{\rm tr} }
\newcommand{\Tr}{{\rm Tr} }
\def\0{{\sst{(0)}}}
\def\1{{\sst{(1)}}}
\def\2{{\sst{(2)}}}
\def\3{{\sst{(3)}}}
\def\4{{\sst{(4)}}}
\def\5{{\sst{(5)}}}
\def\6{{\sst{(6)}}}
\def\7{{\sst{(7)}}}
\def\8{{\sst{(8)}}}
\def\n{{\sst{(n)}}}
\def\cA{{{\cal A}}}
\def\cF{{{\cal F}}}
\def\tV{\widetilde V}
\def\tW{\widetilde W}
\def\tH{\widetilde H}
\def\tE{\widetilde E}
\def\tF{\widetilde F}
\def\tA{\widetilde A}
\def\im{{{\rm i}}}
\def\tY{{{\wtd Y}}}
\def\ep{{\epsilon}}
\def\vep{{\varepsilon}}
\def\R{\rlap{\rm I}\mkern3mu{\rm R}}
\def\bD{{{\bar D}}}

\def\R{\rlap{\rm I}\mkern3mu{\rm R}}
\def\bD{{{\bar D}}}
\def\R{{{\Bbb R}}}
\def\C{{{\Bbb C}}}
\def\H{{{\Bbb H}}}
\def\CP{{{\Bbb C}{\Bbb P}}}
\def\RP{{{\Bbb R}{\Bbb P}}}
\def\Z{{{\Bbb Z}}}
\def\bA{{{\Bbb A}}}
\def\bB{{{\Bbb B}}}
\def\bC{{{\Bbb C}}}
\def\bD{{{\Bbb D}}}
\def\bE{{{\Bbb E}}}
\def\bZ{{{\Bbb Z}}}
\def\Re{{{\frak{Re}}}}
\def\Im{{{\frak{Im}}}}
\def\cosec{{\,\hbox{cosec}\,}}
\def\Gm{{\Gamma_{\!\! -}}}
\def\Gp{{\Gamma_{\!\! +}}}
\def\stan{{standard }}
\def\nonstan{{supernumerary }}

\newcommand{\tamphys}{\it Center for Theoretical Physics,
Texas A\&M University, College Station, TX 77843, USA}
\newcommand{\umich}{\it Michigan Center for Theoretical Physics,
University of Michigan\\ Ann Arbor, MI 48109, USA}
\newcommand{\upenn}{\it Department of Physics and Astronomy,
University of Pennsylvania\\ Philadelphia,  PA 19104, USA}
\newcommand{\SISSA}{\it  SISSA-ISAS and INFN, Sezione di Trieste\\
Via Beirut 2-4, I-34013, Trieste, Italy}

\newcommand{\ihp}{\it Institut Henri Poincar\'e\\
  11 rue Pierre et Marie Curie, F 75231 Paris Cedex 05}

\newcommand{\damtp}{\it DAMTP, Centre for Mathematical Sciences,
 Cambridge University\\ Wilberforce Road, Cambridge CB3 OWA, UK}
\newcommand{\itp}{\it Institute for Theoretical Physics, University of
California\\ Santa Barbara, CA 93106, USA}

\newcommand{\auth}{M. Cveti\v{c}\hoch{\dagger},
H. L\"u\hoch{\star} and C.N. Pope\hoch{\ddagger}}

\begin{document}
\begin{flushright}
\hfill{CTP TAMU-04/02}\ \ \ {UPR-982-T}\ \ \
{MCTP-02-15}\\
{March 2002}\ \ \
{hep-th/0203082}
\end{flushright}


\begin{center}
{ \large {\Large\bf Penrose Limits, PP-Waves and Deformed M2-branes}}

\vspace{20pt}
\auth

\vspace{3pt}
{\hoch{\dagger}\upenn}

\vspace{3pt}


\vspace{3pt}
{\hoch{\star}\umich}

\vspace{3pt}
{\hoch{\ddagger}\tamphys}

\vspace{3pt}

\underline{ABSTRACT}
\end{center}

   Motivated by the recent discussions of the Penrose limit of
AdS$_5\times S^5$, we examine a more general class of supersymmetric
pp-wave solutions of the type IIB theory, with a larger number of
non-vanishing structures in the self-dual 5-form. One of the pp-wave
solutions can be obtained as a Penrose limit of a D3/D3 intersection.
In addition to 16 standard supersymmetries these backgrounds always allow
for supernumerary supersymmetries. The latter are in one-to-one
correspondence with the linearly-realised world-sheet supersymmetries
of the corresponding exactly-solvable type IIB string action. The pp-waves 
provide new examples where supersymmetries will survive in a T-duality
transformation on the $x^+$ coordinate.  The T-dual solutions can be
lifted to give supersymmetric deformed M2-branes in $D=11$.  The
deformed M2-brane is dual to a three-dimensional field theory whose
renormalisation group flow runs from the conformal fixed point in the
infra-red regime to a non-conformal theory as the energy increases.
At a certain intermediate energy scale there is a phase transition
associated with a naked singularity of the M2-brane.  In the
ultra-violet limit the theory is related by T-duality to an
exactly-solvable massive IIB string theory.

{\vfill\leftline{}\vfill
\vskip 5pt
\footnoterule
{\footnotesize \hoch{\dagger} Research supported in part by DOE grant
DE-FG02-95ER40893 and NATO grant 976951. \vskip -12pt} \vskip 14pt
{\footnotesize \hoch{\star} Research supported in full by DOE grant
DE-FG02-95ER40899 \vskip -12pt} \vskip 14pt
{\footnotesize  \hoch{\ddagger} Research supported in part by DOE
grant DE-FG03-95ER40917.\vskip  -12pt}}

\pagebreak
\setcounter{page}{1}


\section{Introduction}\label{introsec}

   Maximally supersymmetric type IIB pp-waves \cite{blafighulpap} turn out to
arise as the Penrose limit \cite{penrose} of AdS$_5\times S^5$
\cite{blafighulpap2,bermalnas}.  The pp-wave provides a background that
gives rise to an exactly-solvable string theory, with free massive
fields in the light-cone gauge \cite{mat,bermalnas}.  This provides a
way of studying the dual ${\cal N}=4$ superconformal field theory in
the Penrose limit \cite{bermalnas}.  Owing to the exact solvability of
the system, there has been a burgeoning activity in the subject
\cite{bur1}--\cite{bur7}.  In this paper we shall consider other
examples of supersymmetric pp-waves, with more general structures for
the 5-form field strength.  Our focus will be on examples where there
are Killing spinors that are independent of the $x^+$ coordinate, 
thereby allowing us to obtain solutions that remain supersymmetric 
after T-dualisation.  After lifting these to $D=11$, we obtain 
supersymmetric deformed M2-branes.   We also show that these pp-waves
correspond to exactly-solvable massive string theories with
linearly-realised world-sheet supersymmetries.

    The Penrose limit of the AdS$_5\times S^5$ solution of the type IIB 
theory is given by
\bea
ds^2 &=& -4 dx^+\, dx^-  - \ft1{16}\mu^2\, z_i^2\, {dx^+}^2 + dz_i^2\,,
\label{55metric}\\
F_5 &=& \mu\, dx^+\wedge \Phi_\4\,,\label{55f5}
\eea
where the 4-form $\Phi_\4$, which is self-dual in the flat metric 
$d z_i^2$ on $\bE^8$, is given by
\be
\Phi_\4 = dz^1\wedge dz^2\wedge dz^3\wedge dz^4 +
dz^5\wedge dz^6\wedge dz^7\wedge dz^8\,.\label{2term}
\ee
This particular solution preserves all the supersymmetry \cite{blafighulpap}.

   This is a special case of a more general class of pp-wave solution,
in which $\Phi_\4$ in (\ref{55f5}) is replaced by any constant
self-dual 4-form on $\bE^8$, and with the metric (\ref{55metric}) generalised 
to
\be
ds^2=  -4 d x^+\, dx^-  + H\, {dx^+}^2 + d z_i^2\,,\label{hmet}
\ee
where $H$ is a function on $\bE^8$ satisfying the equation
\be
\square H = - \ft1{48}\, \mu^2\, |\Phi_\4|^2\,.\label{squareh}
\ee
Different choices of $\Phi_\4$ give rise to different amounts of
preserved supersymmetry; generically there will always be 16 Killing spinors,
with additional ones for special choices $\Phi_\4$, provided also that
$H$ is quadratic in $z^i$, $H=c_0 - \mu_i^2\, z_i^2$, with
appropriate choices for the $\mu_i$.  For example, there are an
additional 16 Killing spinors if $\Phi_\4$ has the form (\ref{2term})
(or a structure related to this one by symmetry) and $H$ is quadratic
with all the $\mu_i$ equal \cite{blafighulpap}.\footnote{Note that 
for the 16 Killing spinors that always occur, one
can include solutions of the homogeneous equation, giving
contributions of the form $Q/(z_i^2)^3$ in $H$.  However for the additional
Killing spinors that arise in special cases, $H$ cannot include such a term.}

   We shall refer to the generic 16 Killing spinors that always occur
in any pp-wave background as ``\stan Killing spinors,'' whilst the
additional ones that occur only in special cases will be referred to
as ``\nonstan Killing spinors.''

    Writing (\ref{hmet}) in the form
\be
ds^2 = H\, (dx^+ -2 H^{-1}\, dx^-)^2 - 4\, H^{-1}\, d{x^-}^2 + 
    d z_i^2\,,\label{prered}
\ee
we can perform a T-duality transformation on the $\del/\del x^+$
Killing direction, thereby obtaining a deformed string solution of the
type IIA theory, where the function $H$ has the 4-form $F_\4=\mu\,
\Phi_\4$ as its source.  Lifting this to $D=11$, one then obtains a
deformed M2-brane solution of M-theory, in which the 4-form carries an 
additional flux $\mu\, \Phi_\4$:
\bea
ds_{11}^2 &=& H^{-2/3}\, (-dt^2 + dx_1^2 +dx_2^2) + H^{1/3}\, d z_i^2\,,
\nn\\
F_\4 &=& dt\wedge dx_1\wedge dx_2\wedge dH^{-1} + \mu\, \Phi_\4\,.
\label{deformedm2}
\eea
The coordinates $t$ and $x_1$ are $2x^-$ and $x^+$ respectively, and
$x_2$ is the eleventh coordinate.  For convenience, we are assuming
here that $H$ is positive so that $x^+$ is spacelike and $x^-$ is
timelike.  This can easily be achieved, for a range of the coordinates
$z^i$, by taking the solution of (\ref{squareh}) to be $H=c_0-
\mu_i^2\, z_i^2$, where $c_0$ is a positive constant.  The inclusion
of this constant is not in contradiction with what one can obtain from
a Penrose limit, since it can be introduced merely by performing a
coordinate transformation $x^-\longrightarrow x^- - \ft14 c_0\, x^+$.
The inclusion of this constant has the merit of making the discussion
of T-duality simpler, since the Killing coordinate can then be taken
to be spacelike rather than timelike.  We shall return to this point
later, in section \ref{tdualsec}.

    Although $\del/\del x^+$ is a Killing direction, allowing the
above T-dualisation to be performed, the situation regarding
supersymmetry is a little more complicated.  For generic choices of
$\Phi_\4$, the associated 16 Killing spinors will all depend
(periodically) on $x^+$, and consequently they will all be projected
out, at the level of the low-energy effective field theory, in the
circle reduction involved in the T-duality transformation.  Of course
at the level of the full string theory, where winding states are
included too, the supersymmetry will survive the T-dualisation
(provided the radius of the circle is appropriately chosen.) This is
an example of the phenomenon of ``supersymmetry without
supersymmetry,'' which was discussed in \cite{susywosusy1,susywosusy2}
(see also \cite{bakas1,bakas2,bakas3} for earlier work on supersymmetry
under T-duality).
The Penrose limit of AdS$_5\times S^5$, for which $\Phi_\4$ is given
by (\ref{2term}), is an example of a special case since there are
\nonstan Killing spinors (16 in this example).  We shall discuss
these special cases below, after first giving a discussion applicable
to the generic situation with only the 16 \stan Killing spinors.

   Deformed M2-branes of the form (\ref{deformedm2}), with the transverse
8-metric $dz_i^2$ replaced by a Ricci-flat metric of special holonomy, 
and $\Phi_\4$ an $L^2$-normalisable self-dual harmonic 4-form, 
have been discussed extensively 
\cite{hawtay,clptrans,kbec,cglpsten,cglphyper,newspin7}.
The condition for a Killing spinor $\ep$ of the
undeformed solution to survive the deformation is that 
\cite{2bec,hawtay,kbec}
\be
\Phi_{abcd}\, \Gamma^{bcd}\, \ep=0\label{susycon}
\ee
in the transverse space.  In our present context we instead take the
8-space to be flat, and $\Phi_\4$ to be constant (and hence
non-normalisable). 

   A solution to (\ref{squareh}) that is sufficient for our 
purposes is
\be
H= c_0 + \fft{Q}{r^6} - \mu_i^2\, z_i^2\,,\label{isotropic}
\ee
where $r^2\equiv z_i^2$, $Q$ is the M2-brane charge, and 
\be
\sum_i \mu_i^2 = \ft1{96} \mu^2\, |\Phi_\4|^2\,.\label{mui}
\ee
We shall refer to $H$ as being ``quadratic in the $z^i$'' if the
charge $Q$ vanishes, regardless of whether or not the constant $c_0$
vanishes.  The criterion for a Killing spinor in the original
undeformed solution (with $Q\ne0$ but with $\mu=0$, and hence $\ft12$
supersymmetry) to remain a Killing spinor after the deformation
($\mu\ne0$) is still (\ref{susycon}), and so by appropriate choices
for the non-vanishing components of $\Phi_\4$ one can arrange for some
fraction of the supersymmetry to be preserved.  

    Let us now turn to the discussion of the \nonstan supersymmetries
that can arise for special $\Phi_\4$ and $\mu_i$ distributions (with
$Q=0$).  As we mentioned, an example of this is the Penrose limit of
AdS$_5\times S^5$, where it was shown in \cite{blafighulpap} that for
$\Phi_\4$ given by (\ref{2term}), and $Q=0$ and the $\mu_i$ all equal
in (\ref{isotropic}), there are 16 \nonstan Killing spinors.  In
section \ref{gensusysec}, we shall show that these are all independent
of $x^+$, implying that they will survive in a T-dualisation, giving a
supersymmetric deformed M2-brane.  However, by naively applying
the criterion (\ref{susycon}) to the expression for $\Phi_\4$ given in
(\ref{2term}), one would draw the conclusion that this particular
M2-brane should have no supersymmetry, since the operator in (\ref{susycon})
has no zero eigenvalues.  As we shall discuss in section
\ref{newm2sec}, in this particular case it is incorrect to check the
supersymmetry by first finding the Killing spinors of the undeformed
solution and then testing to see whether they survive in the criterion
(\ref{susycon}).  Specifically, there are in fact 16 Killing spinors
when $\mu\ne0$ and $Q=0$, but they are disjoint from the 16 that one
has when $Q\ne0$ and $\mu=0$.  Their existence depends on non-zero
contributions from the $\Phi_\4$ terms in the supersymmetry
transformation rules cancelling against the other terms that are
present even in the undeformed background.  Thus although the general
solution with both $Q$ and $\mu$ non-vanishing has no supersymmetry in
this example, one gets two disjoint sets of 16 Killing spinors in the
two cases $Q=0$ or $\mu=0$.  

    In fact the Penrose limit of AdS$_5\times S^5$ is just one of many
possible examples of special cases with extra $x^+$-independent
supersymmetries, which, in the M2-brane picture, cannot be found by 
simply applying the criterion (\ref{susycon}).  In summary, the 
supersymmetry criterion (\ref{susycon}) is correct for testing ``\stan
Killing spinors,'' but not for testing ``\nonstan Killing spinors,'' 
in the T-dualised M-theory picture. We shall discuss these issues further 
in section \ref{xpsec}.
    
   One of the purposes of this paper is to consider various possibilities
for the constant self-dual 4-form $\Phi_\4$ in (\ref{55f5}), focusing
on those cases where the corresponding T-dualised deformed solutions
have surviving supersymmetries.  In general it is not clear that the
type IIB pp-wave solutions with $\Phi_\4$ more general than
(\ref{2term}) will have interpretations as Penrose limits.  However,
in one example we find that such an interpretation does arise.  From
the deformed M2-brane point of view, this example is inspired by
considering the transverse space $\bE^8$ as an orbifold limit of
K3$\times$K3, where the harmonic 4-form is a direct product of the
K\"ahler forms of the two K3 factors;
\be
\Phi_\4 = (dz^1\wedge dz^2+dz^3\wedge dz^4)\wedge 
         (dz^5\wedge dz^6 + dz^7\wedge dz^8)\,.\label{k3k3}
\ee
It is straightforward to verify from (\ref{susycon}) that the
associated deformed M2-brane will have $\ft14$ supersymmetry (\ie 8
Killing spinors), if $Q\ne0$.  When $Q=0$ we find an additional 8
\nonstan Killing spinors, if four of the $\mu_i$ in (\ref{isotropic})
vanish, and the other four are equal and non-zero. We shall show in
section \ref{d3d3sec} that from the type IIB point of view, this
half-supersymmetric pp-wave solution can be obtained as a Penrose
limit of the near-horizon limit of an intersection of two D3-branes.
This near-horizon limit itself is AdS$_3\times S^3\times T^4$ (or
AdS$_3\times S^3\times$K3).

   In section \ref{gensusysec} we discuss the supersymmetry of the
general pp-wave solutions, focusing in particular on the question of
whether there are Killing spinors that are independent of the $x^+$
coordinate, and thus will survive at the field theory level after a
T-duality transformation.  In section \ref{xpsec}, we present a general
class of $\Phi_\4$ structures that will give rise to $x^+$-independent
Killing spinors.  We obtain examples that give rise to deformed M2-branes
that preserve a variety of fractions of supersymmetry.  For example, if
$Q$ is non-zero (implying that only the 16 \stan supersymmetries
in the type IIB pp-wave arise), we can obtain supersymmetry fractions
$n/16$ in the associated deformed M2-branes, with $1\le n\le 6$.
Even more possibilities arise when the special cases where the pp-waves
have \nonstan supersymmetries are considered.  
In section \ref{actionsec} we consider the solvable massive string actions
associated with these pp-wave solutions.  

   In section \ref{tdualsec} we address the issue of the spacelike or
timelike nature of the $x^+$ coordinate, and its effect on the nature
of the T-dualised theories.  In section \ref{newm2sec} we study the
supersymmetry of the deformed M2-branes that are T-dual to the
pp-waves, paying particular attention to the cases where there are
\nonstan Killing spinors.  As an illustrative example, we examine in
detail the example of the Penrose limit of AdS$_5\times S^5$, showing
how it is supersymmetric despite violating the usual supersymmetry
criterion (\ref{susycon}).

\section{D3/D3 brane intersection and its Penrose limit}\label{d3d3sec}

    Let us begin by considering a standard D3/D3 brane intersection, for
which the metric and self-dual 5-form are given by
\bea
ds_{10}^2 &=& (H_1\, H_2)^{-1/2}\, \Big[ -dt^2 + dx_1^2 + H_1\, 
      (dz_1^2 + dz_2^2) + H_2\, (dz_3^2 + dz_4^2)\nn\\
&& + H_1\, H_2\, (dr^2 + r^2\, d\Omega_3^2)\Big]\,,\nn\\
F_\5 &=& -dt\wedge dx_1\wedge (dz_1\wedge dz_2 + dz_3\wedge dz_4)\wedge 
   d(H_1^{-1} + H_2^{-1}) + \hbox{dual}\,,
\eea
where
\be
H_1 = 1 + \fft{Q_1}{r^2}\,,\qquad H_2=1 + \fft{Q_2}{r^2}\,.
\ee
For simplicity, we shall take $Q_1=Q_2=\lambda^2$.  In the near-horizon 
limit, we then find that the metric becomes
\be
ds_{10}^2 = \lambda^2\, \Big[ d\Omega_3^2 + ds_3^2\Big] + dz_1^2
+ dz_2^2 + dz_3^2 + dz_4^2 \,.
\ee
After transforming from the Poincar\'e coordinates $(r,t,x_1)$ to global
coordinates $(\rho, t,\gamma)$ in the AdS$_3$ metric $ds_3^2$, and
writing
\be
d\Omega_3^2 = d\theta^2 + \cos^2\theta\, d\psi^2 + \sin^2\theta\, d\phi^2
\,,\qquad ds_3^2= d\rho^2 -\cosh^2\rho\, dt^2 + \sinh^2\rho\, d\gamma^2\,,
\ee
the Penrose limit can be taken as
\be
\rho\longrightarrow \fft{\rho}{\lambda}\,,\qquad
\theta\longrightarrow \fft{\theta}{\lambda}\,,\qquad 
t= x^+ +\fft{x^-}{\lambda^2}\,,\qquad \psi= x^+ -\fft{x^-}{\lambda^2}\,,
\label{limit}
\ee
with the constant $\lambda$ sent to infinity.  By this means, the
solution becomes the pp-wave
\bea
ds_{10}^2 &=& - 4 dx^+\, dx^-  + H\, d{x^+}^2 + d z_i^2 
\,,\nn\\
F_\5 &=& \mu\, dx^+\wedge (dz_1\wedge dz_2 + dz_3\wedge dz_4)\wedge
    (dz_5\wedge dz_6 + dz_7\wedge dz_8)\,,\label{penroselim}
\eea
where $z_5=\theta\, \cos\phi$, $z_6= \theta\, 
\sin\phi$, $z_7=\rho\, \cos\gamma$, $z_8=\rho\, \sin\gamma$. 
(The constant $\mu$ is introduced here by making the replacements $x^+
\longrightarrow \mu\, x^+$, $x^-\longrightarrow x^-/\mu$.)
The function $H$ is given by
\be
H = - \ft14 \mu^2\, \sum_{i=5}^8 z_i^2\,.\label{h5678}
\ee
Comparing with (\ref{55f5}), we see that the self-dual 4-form $\Phi_\4$ is 
precisely of the form given in (\ref{k3k3}).  

   It is worth pointing out that in this example, and indeed in 
all analogous Penrose limits, we can generalise (\ref{limit}) in the
following way.  The expressions for $t$ and $\psi$ in (\ref{limit}) can
be replaced by
\be
t= (1-\ft14 c_0\, \lambda^{-2})\, x^+ + \fft{x^-}{\lambda^2}\,,\qquad
\psi=  (1-\ft14 c_0\, \lambda^{-2})\, x^+ -  \fft{x^-}{\lambda^2}\,,
\ee
where $c_0$ is a constant.  Now we obtain a metric of the same form as
in (\ref{penroselim}), except that now $H$ is given by
\be
H= c_0 - \ft14 \mu^2\, \sum_{i=5}^8 z_i^2\,.
\ee
This means that there can be a regime where $H$ is positive, implying
that $x^+$ is then a spatial coordinate and $x^-$ a timelike
coordinate.  Note that this same generalisation can be made in the
standard Penrose limit of AdS$_5\times S^5$, allowing $H$ in equation
(\ref{hmet}) to become $H=c_0-\ft1{16} \mu^2\, z_i^2$.  In fact the
generalised Penrose limit that we are introducing here is nothing but
a general coordinate transformation in which one sends
$x^-\longrightarrow x^- - \ft14 c_0\, x^+$.

   Note that although the summation range is restricted in
(\ref{h5678}), we could instead solve the type IIB equations of motion
for configurations of the form (\ref{penroselim}) with a summation
over the entire range $1\le i \le 8$, although then the interpretation
as a Penrose limit of the D3/D3 brane intersection would be lost.
Furthermore, as we shall see in the next section, one would have only
16 rather than 24 Killing spinors.

   In order to study the supersymmetry of these type IIB solutions, 
it will prove to be useful to give a more general discussion of 
supersymmetry for solutions of the form (\ref{hmet}) and (\ref{55f5}).
This forms the topic of the next section.

\section{Supersymmetry analysis for the pp-wave solutions}\label{gensusysec}

   The discussion in this section follows the strategy described in
\cite{blafighulpap}, with appropriate generalisation and adjustment to
our notation and conventions.  The spinor covariant derivative in the
metric (\ref{hmet}) can be seen to take the form
\be
\nabla_+= \del_+ + \ft14 \del_i H\, \Gm\, \Gamma_i\,,\qquad
\nabla_-=\del_-\,,\qquad \nabla_i=\del_i\,.
\ee
With a suitable normalisation for the self-dual 5-form, the supercovariant
derivative $D_M$ is given by
\be
D_M = \nabla_M + \im\,\, \Omega_M\,,
\ee
where
\be
\Omega_M= \ft1{192} F_{MN_1\cdots N_4}\, \Gamma^{N_1\cdots N_4}\,.
\ee
With $F_\5$ given by (\ref{55f5}), one then has
\be
\Omega_-=0\,,\qquad \Omega_+ = \ft18 \mu\, W\,,
\qquad \Omega_i = -\ft1{16} \mu\, \Gm\, [\Gamma_i,W]\,,
\ee
where
\be
W\equiv \ft1{24} \Phi_{ijk\ell}\, \Gamma_{ijk\ell}\,.\label{wdef}
\ee

  Note that in the type IIB theory there are two independent 
supersymmetry parameters $\ep$, which are both of the same chirality.
Thus each could potentially give up to 16 Killing spinors, implying a 
total maximum of 32.

    Following the arguments presented in \cite{blafighulpap}, one can
now straightforwardly show that the Killing spinors, 
which satisfy $D_M\,\ep=0$,
are independent of $x^-$, and are given by
\be
\ep= (1-\im\, z^i\, \Omega_i)\, \chi\,,\label{zdep}
\ee
where $\chi$ has only $x^+$ dependence, governed by
\be
\del_+\, \chi + \im\, \mu\, W\, \chi=0\,.\label{xplus}
\ee
Additionally, one has the requirement
$\mu\, z^i\,[W,\Omega_i]\, \chi + \ft14 \del_i H\, \Gm\, \Gamma_i\, 
\chi=0$, which therefore implies, using the fact that for the chiral 
spinors $\chi$ we have $W\, \Gm\, \chi=0$,
\be
(\mu^2\, z^i\, W^2 + 32 \del_i H)\, \Gamma_i\, \Gm\, \chi=0
\,.\label{susyfrac}
\ee
It is this equation that determines the number of Killing spinors.
Note that for any solution $H(z_i)$ one is guaranteed to have at least
16 Killing spinors $\chi= \Gm\, \chi_0$, since every term in
(\ref{susyfrac}) contains a factor of $\Gm$.  It follows from
(\ref{zdep}) that these 16 Killing spinors are independent of the
$z^i$ coordinates.  These are the \stan Killing spinors that we
spoke of earlier.

   If $H$ is quadratic (\ie $Q=0$ in (\ref{isotropic})), so that
$\del_i H= -2\mu_i^2\, z^i$, then the possibility exists for specific
$\mu_i$ and $\Phi_\4$ that there may be further solutions to
(\ref{susyfrac}).  These \nonstan Killing spinors are constructed
using spinors $\chi$ that satisfy 
\be
\Gp\, \chi=0\,,\qquad 
(\mu^2\, z^i\, W^2 + 32 \del_i H)\, \Gamma_i\, \chi=0
\,.\label{susyfrac2}
\ee
Note that because the chiral spinors $\chi$ satisfy $\G_+\, \chi=0$,
we have $W\, \chi=0$ for any $W$ given by (\ref{wdef}) for a self-dual
$\Phi_\4$:
\be
\Gamma_{11}\, \chi=\chi \quad \hbox{and}\quad \Gp\, \chi=0 \quad
\Longrightarrow\quad  W\, \chi=0 \,.\label{wchi}
\ee
It then follows from (\ref{xpdep}) that
all \nonstan Killing spinors are necessarily independent of $x^+$.

   Before discussing our new pp-wave solutions it is useful to review
in our notation the construction of the 16 \nonstan Killing spinors
in the Penrose limit of AdS$_5\times S^5$, which were found in
\cite{blafighulpap}.  The 4-form $\Phi_\4$ is given by
(\ref{2term}), and hence from (\ref{wdef}) we have
\be
W=\Gamma_{1234} + \Gamma_{5678}\,.\label{2termw}
\ee
It can be seen that we therefore have $W^2\, \Gamma_i\, \chi=
4\Gamma_i\, \chi$ and so the maximum of 32 Killing-spinor solutions to
(\ref{susyfrac}) is achieved (16 \stan plus 16 \nonstan Killing spinors),
provided that one has
\be
H= c_0 - \ft1{16} \mu^2\, z_i^2\,.
\ee
The constant $c_0$ was not included in the discussion in
\cite{blafighulpap}, nor is it usually presented in the Penrose limit
of AdS$_5\times S^5$, but its inclusion does not affect the
conclusions about supersymmetry, since $H$ appears in the spin
connection only via its derivative. Furthermore, as we showed in section
\ref{d3d3sec}, it can easily be introduced in the Penrose limit, via 
a coordinate transformation $x^-\longrightarrow x^- -\ft14 c_0\, x^+$.

  Turning now to our pp-wave solution in section \ref{d3d3sec},
associated with the Penrose limit of AdS$_3\times S^3$, we shall have
\be
W = \Gamma_{1234} + \Gamma_{5678} +\Gamma_{1278}+
\Gamma_{3456}\,.\label{4is}
\ee
There are, as always, 16 \stan Killing spinors, corresponding to
$\chi= \Gm\, \chi_0$.  After some algebra, we find that there are in
addition 8 \nonstan Killing spinors satisfying (\ref{susyfrac2}), with
$W^2\, \Gamma_i\, \chi = 0$ for $i=1,2,3, 4$ and $W^2\, \Gamma_i\, \chi=
16\Gamma_i\, \chi$ for $i=5,6,7,8$ (in a convenient labelling convention 
for the gamma matrices).  Thus we obtain the \nonstan 8 Killing 
spinors if $H$ is given by
\be
H= c_0 - \ft14 \mu^2 \sum_{i=5}^8 z_i^2\,.
\ee
Thus the solution preserves $\ft12 + \ft14=\ft34$ of the supersymmetry.
If any other distribution of $\mu_i$ (besides relabelling) is chosen
(still satisfying the field equation (\ref{mui})), all the 8 \nonstan 
Killing spinors will be lost.

   Having found all the Killing spinors, we can now address the question
of which are independent of $x^+$; this is determined by
(\ref{xplus}), whose solution is 
\be
\chi = e^{-\im\, \fft18 \mu\, x^+\, W}\, \eta\,.\label{xpdep}
\ee
If a particular solution for $\chi$ has the property that $W\,
\chi=0$, then the associated Killing spinor will be independent of the
coordinate $x^+$.  In particular, from (\ref{wchi}), this means that all 
\nonstan Killing spinors, coming from solutions of
(\ref{susyfrac2}), will be independent of $x^+$.  For the \stan 16
Killing spinors, on the other hand, the requirement $W\, \chi=0$ is a
further condition, which may or may not have solutions, depending on
the structure of $W$.  

   For our new case where $W$ is given by (\ref{4is}), it is
straightforward to see that 8 of the 16 \stan Killing spinors are
annihilated by the operator $W$, and thus these 8 do not depend on the
coordinate $x^+$.  In total, therefore, we have the 8 \nonstan
Killing spinors and 8 of the \stan Killing spinors that do not
depend on $x^+$.  These 16 Killing spinors will therefore survive
under a T-duality transformation, and so after lifting to $D=11$ we
should obtain a deformed M2-brane that preserves $\ft12$ the
supersymmetry.  The 8 that are \stan Killing spinors, which, as we
saw earlier, are also independent of $z^i$, are the ones that one
would expect in $D=11$, by simply applying the standard supersymmetry
condition (\ref{susycon}).  The 8 coming from the \nonstan
Killing spinors in the pp-wave are rather remarkable from an M-theory
viewpoint; we shall discuss these in detail in section \ref{newm2sec}.

   In the case of the Penrose limit of AdS$_5\times S^5$, $W$ given in
(\ref{2termw}) does not annihilate any of the 16 \stan Killing
spinors, but it does, as we saw earlier, annihilate the
\nonstan Killing spinors, and so these 16 are independent of
$x^+$.  This verifies a statement made in \cite{bermalnas}.  It
contradicts a statement in \cite{blafighulpap}, where it was stated
that all 32 Killing spinors depend on the $x^+$ coordinate (called
$x^-$ in \cite{blafighulpap}).  The explanation for this is that
\cite{blafighulpap} did not take into account that $W$ has 16 zero
eigenvalues.\footnote{Specifically, the 16 Killing spinors annihilated
by $\Gm$ are not annihilated by $W$, and so they depend on $x^+$ but
not on $z^i$.  The remaining 16 Killing spinors are independent of
$x^+$ but they do depend on the $z^i$.}

    The fact that this Penrose limit of AdS$_5\times S^5$ has 16
Killing spinors that are independent of $x^+$ again has a remarkable
consequence in this case; that there should exist, in the T-dualised
picture, a deformed M2-brane whose additional 4-form flux violates the
usual supersymmetry condition (\ref{susycon}) and yet in fact
preserves $\ft12$ of the supersymmetry.  We shall return to this
in section \ref{newm2sec}, where we shall construct this
M2-brane solution explicitly, and demonstrate its supersymmetry.

\section{$x^+$-independent Killing spinors in pp-waves}\label{xpsec}

   Having seen in the previous section that there can exist pp-waves
with Killing spinors that do not depend on the coordinate $x^+$, it
becomes of interest to classify the possible 5-form structures and
the distributions of $\mu_i$ coefficients in (\ref{isotropic}), to see
what fractions of $x^+$-independent Killing spinors can be achieved.

\subsection{$x^+$-independent \stan Killing spinors}

    We shall first study this question for the 16 \stan Killing
spinors, which are all annihilated by $\Gm$ (and thus are all independent 
of the $z^i$ coordinates).  These exist for any 
choice of $H$, provided only that the field equation (\ref{squareh}) is
satisfied.  In terms of $W$, defined in (\ref{wdef}) this is
\be
\square H = -\ft1{64}\, \mu^2\, \tr W^2\,.\label{squareh2}
\ee
It remains, therefore, to check how many of the 16 \stan Killing
spinors are annihilated by $W$ since as we showed from (\ref{xpdep}),
this is the condition for them to be independent of $x^+$.  In turn,
this subset of the 16 \stan Killing spinors will survive at the
field theory level in a T-duality transformation.

  As we discussed earlier, the T-dualised solutions can be lifted to
$D=11$, where they become deformed M2-branes with an extra self-dual
4-form flux in the 8-dimensional flat transverse space.  One might
expect that the smallest degree of unbroken supersymmetry that could 
be achieved for such deformed M2-branes would be in a case where
the 4-form flux was related to that seen in the harmonic 4-form of a
transverse space of Spin(7) holonomy.  Motivated by this, it is therefore
natural here to consider a self-dual 4-form $\Phi_\4= L_\4 + {*L_\4}$ 
where $L_\4$ has 7 structures, which can be taken to be
\be
L_\4= m_1\, dz^{1234} + m_2\, dz^{1256} + m_3\, dz^{1357} + 
m_4 \, dz^{1467} + m_5\, dz^{2367} + m_6\, dz^{2457} + m_7\, 
dz^{3456}\,,\label{ledf}
\ee
where $dz^{ijk\ell}\equiv dz^i\wedge dz^j\wedge dz^k\wedge dz^\ell$, and
the $m_i$ are constants.  From this, one constructs the matrix $W$ 
given in (\ref{wdef}), which can be written as 
\be
W = \sum_{\a=1}^7 m_\a\, W_\a\,,\label{wadef}
\ee
and then one determines what fraction of the 16 \stan Killing spinors
(which satisfy $\chi=\Gm\, \chi_0$) are annihilated by $W$.  In fact
the choice of seven structures in (\ref{ledf}) can be characterised by
the fact that they give the maximal set of $W_\a$ that all commute.

   It is a simple exercise to obtain the eigenvalues of the matrix
$W$, since the terms $W_\a$ all commute.
Projected into the 16-dimensional subspace of chiral spinors $\chi$ 
satisfying $\Gm\, \chi=0$ (\ie 2 copies of an 8-dimensional subspace
of the $32\times 32$ matrix $W$, corresponding to the ${\cal N}=2$
supersymmetry in the type IIB theory), we find that they
are given by $\lambda_i$, where 
\bea
\lambda_1 &=& 2(m_1 + m_2 - m_3 + m_4 - m_5 - m_6 - m_7)\,,\nn\\
\lambda_2 &=& -2(m_1 - m_2 + m_3 + m_4 + m_5 - m_6 - m_7)\,,\nn\\
\lambda_3 &=& -2(m_1 - m_2 - m_3 - m_4 - m_5 + m_6 - m_7)\,,\nn\\
\lambda_4 &=& 2(m_1 + m_2 + m_3 - m_4 + m_5 + m_6 - m_7)\,,\nn\\
\lambda_5 &=& -2(m_1 + m_2 + m_3 - m_4 - m_5 - m_6 + m_7)\,,\nn\\
\lambda_6 &=& 2(m_1 - m_2 - m_3 - m_4 + m_5 - m_6 + m_7)\,,\nn\\
\lambda_7 &=& 2(m_1 - m_2 + m_3 + m_4 - m_5 + m_6 + m_7)\,,\nn\\
\lambda_8 &=& -2(m_1 + m_2 - m_3 + m_4 + m_5 + m_6 + m_7)\,,\label{meqs}
\eea
If the constants $m_\a$ are chosen so that any number $n\le6$ of the
$\lambda_i$ vanish, there will then be $2n$ \stan Killing
spinors in the type IIB solution that are independent of $x^+$.  If 7
or 8 of the $\lambda_i$ vanish, then all the $m_\a$ vanish and the
solution becomes trivial.

    The \stan Killing spinors that we have been considering in this
subsection are all annihilated by $\Gm$, and so from (\ref{zdep}) they
are all independent of $z^i$.  Thus the subsets that are annihilated
by $W$ are independent of all the coordinates.  

\subsection{$x^+$-independent \nonstan Killing spinors}

   As we showed in section \ref{gensusysec}, all the \nonstan
Killing spinors, satisfying (\ref{susyfrac2}), are independent of
$x^+$, and so it is merely necessary to count them.  
For these Killing spinors, the charge $Q$ in (\ref{isotropic}) must
vanish, so that $H$ is quadratic in the $z^i$.  Furthermore, we find
that we must have
\be
\mu_i^2= \ft1{64} \mu^2\, \lambda_i^2\,,\label{muicon}
\ee
and that the number of these \nonstan Killing spinors is equal to
the degeneracy $k$ of the least degenerate of the squared eigenvalues
$\lambda_i^2$, multiplied by a factor of 2 because of the ${\cal N}=2$
supersymmetry of the type IIB theory.  

    To prove these facts, we begin by noting that there exists a similarity
transformation matrix $S$ such that $S\, W\, S^t$ is diagonal. The
$32\times 32$ matrix itself has 16 zero eigenvalues and 2 of each of the 
$\lambda_i$ given in (\ref{meqs}).  We can therefore change to a new basis
for the gamma matrices, $\Gamma_A\longrightarrow S\, \Gamma_A\, S^t$, in
which $W$ is diagonal. After projection into the subspace
of chiral spinors that are also annihilated by $\Gp$, we have just the eight 
diagonal entries $\lambda_i$.  

   In this diagonal basis, we find that for each $i$ and $j$ in the range 
1 to 8, the matrix $-\Gamma_{ij}\, W\, \Gamma_{ij}$ is again diagonal, with
entries that are a permutation of those in $W$.  Furthermore, for
each $W_\a$ in (\ref{wadef}) we have $\Gamma_{ij}\, W_\a\, \Gamma_{ij} = 
\pm W_\a$, with a sign that depends on $i$ and $j$.    Suppose now that
$\Gamma_1\, \chi$ is an eigenvector of $W$ with eigenvalue
$\lambda$.  It follows, by acting with $\Gamma_{1j}$, that $\Gamma_j\, \chi$
will be an eigenvector of the corresponding permuted matrix, say $\wtd W$,
with the same eigenvalue.   However, the permuted matrix can be transformed
back into $W$ itself by an appropriate set of sign reversals for the
$m_\a$.  It follows that $\Gamma_j\, \chi$ is therefore an eigenvector of
$W$ with one of the other eigenvalues of $W$.  Thus we have established
that the set of eigenvectors of $W$ can be expressed as $\Gamma_i\, \chi$,
once one has established that any one of these is one of the eigenvectors,
and so with a suitable labelling order for the gamma matrices we have
\be
W\, \Gamma_i\, \chi = \lambda_i\, \Gamma_i\, \chi\,,\qquad 1\le i\le 8\,.
\label{wevs}
\ee
Substituting into (\ref{susyfrac2}), with $H$ given by (\ref{isotropic})
and $Q=0$, we therefore obtain the result (\ref{muicon}).
    
   By this means we can obtain 8 solutions to the conditions 
(\ref{susyfrac2}).  However, in general each solution will require that
the set of constants $\mu_i$ in $H$ will be permuted differently from 
each other solution.  Thus for a fixed choice for the $\mu_i$, there
will in general only be one solution to (\ref{susyfrac2}), implying, after
the doubling because of ${\cal N}=2$ supersymmetry in type IIB, two
\nonstan Killing spinors.  If, however, there are degeneracies among
the $\lambda_i$, there can accordingly be more than one solution to 
(\ref{susyfrac2}) with the {\it same} set of $\mu_i$ in $H$.  
It is clear that the largest number of solutions that one could have
for a given choice of $\mu_i$ is therefore equal to the smallest
degeneracy factor, $k$, among the eigenvalues $\lambda_i$.  It turns out that
this largest number is in fact attained, and so we get $2k$ \nonstan
Killing spinors in the pp-wave solution.

    Consider, for example, the case of the Penrose limit of AdS$_5\times
S^5$, where we have $m_1=1$ and all other $m_\a=0$, implying that all
$\lambda_i^2=4$.  The smallest degeneracy is therefore $k=8$, and we
recover the 16 \nonstan Killing spinors previously found
\cite{blafighulpap} in this example.  For the pp-wave associated with
the Penrose limit of AdS$_3\times S^3$, constructed in section
\ref{d3d3sec}, we have $m_1=m_7=1$, with all other $m_\a$ vanishing.
This implies $\lambda_1^2=\lambda_2^2=\lambda_3^2=\lambda_4^2=0$, and
$\lambda_5^2=\lambda_6^2=\lambda_7^2=\lambda_8^2=16$, and hence $k=4$.
This reproduces the 8 \nonstan Killing spinors that we found for
this example in (\ref{gensusysec}).

   In general, the possible values of least degeneracy that can occur
are $k=1$, 2, 4 or 8, implying 2, 4, 8 or 16 \nonstan Killing spinors. 
It is interesting to note that there must therefore always be at least
two \nonstan Killing spinors for any of the configurations for
$\Phi_\4$ contained within (\ref{ledf}), provided, of course, that the
$\mu_i$ are chosen according to (\ref{muicon}) (and $Q$ in (\ref{isotropic})
is set to zero).

\subsection{Supersymmetry of the deformed M2-branes}

   Having determined the numbers of $x^+$-independent Killing spinors
for these pp-wave solutions of the type IIB theory, we can now directly
examine the associated deformed M2-brane solutions in $D=11$ supergravity.

    In general for a deformed M2-brane with $H$ given by
(\ref{isotropic}), with $Q\ne0$, the criterion for unbroken
supersymmetry is given by substituting the 4-form $\Phi_\4$ in the
transverse space into (\ref{susycon}).  After examining all the cases,
we find that the numbers of Killing spinors obtained by setting
$n\le6$ of the $\lambda_i$ in (\ref{meqs}) to zero is indeed $2n$, as
is expected from T-duality.  It is interesting to note that the
supersymmetry fractions that are thus achieved, namely
$\{\ft1{16},\ft18,\ft3{16},\ft14,\ft5{16},\ft38\}$, include some
unusual values.  The first four are seen also in regular deformed
M2-brane solutions using Spin(7), K\"ahler$_8$, hyper-K\"ahler and
K3$\times T^4$ manifolds respectively for the transverse 8-space.  The
final two examples, with $\ft5{16}$ and $\ft38$ supersymmetry, do not
have any known corresponding regular counterparts.

   When $Q=0$, and the $\mu_i$ are given by (\ref{muicon}), the
existence of the $2k$ \nonstan Killing spinors, which are
$x^+$-independent, in the type IIB pp-wave implies that there should
be $2k$ further Killing spinors in the T-dualised deformed M2-brane.
As we shall show in section \ref{newm2sec}, these Killing spinors do
indeed exist in the deformed M2-branes, but they arise in a somewhat
subtle way, since the usual supersymmetry criterion (\ref{susycon}) is
not satisfied.  This is related to their non-standard $z^i$
dependence, which can be seen in the type IIB picture from equation
(\ref{zdep}).

\section{String actions on the pp-waves}\label{actionsec}

  The light-cone gauge string actions for the pp-wave solutions we
have obtained in this paper are given by
\be
{\cal S} = \fft{1}{2\pi \alpha'}\, \int d\tau
\int_0^{2\pi\a'\, p^+}d\sigma\,  {\cal L}\,,
\ee
where
\be 
{\cal L} = \sum_{i=1}^8
(\ft12 \dot z_i^2 -\ft12 z_i'^2 - \ft1{2} \mu_i^2\, z_i^2)
+ \im \, \Psi(\not\del + \ft18\mu\, W)\,\Gp\, \Psi\,,\label{stringact}
\ee
where $W$ is given by (\ref{wdef}).\footnote{This result can in
general be obtained by light-cone gauge fixing in the type IIB 
Green-Schwarz action derived in \cite{clps}.  The results for 
the cases of Penrose limits for AdS$_5\times S^5$ and AdS$_3\times
S^3$ were obtained in \cite{mat,bermalnas}.}
This is an exactly-solvable massive
free string theory, whose character is determined by $W$ and the pattern of
non-vanishing $\mu_i$.  The number of zero eigenvalues of $W$ associated
with eigenspinors that are not annihilated by $\Gm$ determines the
number of massless fermions.  In fact the fermion masses are given by 
$\ft18\mu\, \lambda_i$, where the $\lambda_i$ are the eigenvalues of
$W$, given for our examples in (\ref{meqs}).  We have presented the
action in the case where the additive constant $c_0$ in the function $H$
is set to zero.  If it is non-zero, one gets an additional additive 
constant in ${\cal L}$; this does not affect the exact solvability.

   In the case associated with the Penrose limit of AdS$_5\times S^5$,
all the relevant eigenvalues of $W$ are non-zero, and thus one has 8
massive fermions (with equal mass).  All the bosonic masses are equal
too, since $H$ is isotropic in $z^i$.  This solution has the
\nonstan 16 Killing spinors that are independent of $x^+$ (the time
coordinate in this context), implying that the linearly-realised
supersymmetry commutes with the Hamiltonian and hence that the
fermionic and bosonic masses are equal here \cite{bermalnas}.

  In our further examples of supersymmetric pp-waves with $\Phi_\4$
given by (\ref{ledf}), the matrix $W$ can have $n$ zero eigenvalues
when $1\le n\le 6$ of the $\lambda_i$ in (\ref{meqs}) vanish.  This
implies that $n$ fermions are massless while $(8-n)$ are massive.  For
a general distribution of $\mu_i$ satisfying (\ref{mui}), the masses
of the bosons and fermions are not equal.  This is because there are
no \nonstan Killing spinors in general, and hence there are no
linearly-realised supersymmetries that would imply a bose/fermi mass
equality. However, if instead we choose the $\mu_i$ to be given by
(\ref{muicon}), the masses of the bosons will match with the masses of
the fermions, suggesting the existence of linearly-realised
supersymmetries.  Indeed, as we showed, \nonstan Killing spinors
arise in this case, which give rise to these linearly-realised
supersymmetries.  The number of these \nonstan Killing spinors is
governed by the smallest degeneracy $k$ among the eigenvalues
$\lambda_i$ of $W$, which are the masses of the fermions.  The larger
the degeneracy, the larger the number of equal-mass fields.  This is
consistent with the associated larger number of linearly-realised
supersymmetries.

   The example of the Penrose limit of the D3/D3 system,
discussed in section \ref{d3d3sec}, is a special case of (\ref{ledf}) 
where $\Phi_\4$ is given by (\ref{k3k3}).  The light-cone Lagrangian is
given by
\be 
{\cal L} = \sum_{i=1}^8
(\ft12 \dot z_i^2 -\ft12 z_i'^2) - \ft14 \mu^2\, \sum_{i=5}^8 z_i^2
+ \im \, \Psi(\not\del + \ft18 \mu\, W)\,\Gp\, \Psi\,,
\ee
with $W$ given by (\ref{4is}).  In this case there are four massive
fermions and bosons (with equal masses), and four massless fermions
and bosons.  The equality of the fermion and boson masses is a consequence
of having the \nonstan Killing spinors that we discussed earlier.
This string action is similar to the result one obtains
from the D1/D5 system \cite{bermalnas,bur4}.  However, the fermion
structure is different because in the D1/D5 system there is a
non-vanishing R-R 3-form, whilst in our case it is the 5-form that is
non-vanishing.  Note that we can replace the 4-space whose coordinates
are $z^i$ for $1\le i\le 4$ by a $T^4/Z_N$ orbifold.  The twisted
states arising from this orbifolding can be analysed in a conventional
way, since the associated target-space coordinates correspond to
massless free fields.

\section{Spacelike vs.~timelike T-duality, and  
phase transitions}\label{tdualsec}

     In presenting the pp-wave solution (\ref{hmet}) or
(\ref{prered}), we have chosen to treat $x^+$ as a spacelike
coordinate, by taking $H$ to be positive.  This means that we can
perform a spacelike T-duality transformation on the coordinate $x^+$,
leading to deformed M2-brane solutions in M-theory.  As we discussed
in section \ref{introsec}, the positivity of $H$ can be justified
since solutions to (\ref{squareh}) can be taken to be
(\ref{isotropic}), implying positivity in some region of spacetime.
Moreover, even though the term involving $Q$ would not arise in a
Penrose limit the constant $c_0$ does, and so it is always possible to
find a region of spacetime where $x^+$ is spacelike.

     On the other hand, the pp-wave solution that is of particular
interest in the context of Penrose limits corresponds to setting
$Q=0$.  For sufficiently large values of $z_i^2$ the coordinate $x^+$
becomes timelike, and indeed in the discussion of solvable string
actions one chooses the light-cone gauge where $x^+$ is set equal to
the worldsheet time coordinate \cite{mat,bermalnas}.  Whilst T-duality
generalises to a timelike $U(1)$ isometry in the heterotic string
theory \cite{moore}, it breaks down for the type II string theories
because of the Ramond-Ramond sector \cite{cllpst}.  In fact it has
been proposed that the type IIA and IIB theories are timelike T-dual
to new theories called type IIB$^*$ and type IIA$^*$ respectively,
where all the R-R field strengths have kinetic terms of reversed sign
\cite{hulltime1}.  Thus an alternative viewpoint is to consider a
timelike T-duality on the $x^+$ direction for the pp-wave solution, in
the region where $H<0$, and obtain now a regular 2-brane in the M$^*$
theory in $(2,9)$ spacetime signature that was introduced in
\cite{hulltime2}.  

   Whichever view one takes, it will not affect the conclusions 
about the supersymmetry of the T-dualisations of the
pp-wave solutions that we have obtained
here.  From a practical point of view, it is simpler to analyse the
supersymmetry in the region where $H$ is positive.
   
    If we consider the deformed M2-branes with $H$ given by
(\ref{isotropic}) where $Q$ as well as $\mu$ is non-zero, the solution
close to $r=0$ is the near-horizon limit AdS$_4\times S^7$, which is
dual to a three-dimensional superconformal field theory.  On the other
hand at large $r$, the solution approaches a deformed 2-brane of
M$^*$-theory that is T-dual to a pp-wave in the type IIB string.  Thus
the naked singularity of the M2-brane that arises as $H$ passes
through 0 can be argued to be a low-energy artefact of
eleven-dimensional supergravity.  It is in fact the point where a
phase transition would take place, as one passes from the M-theory to
M$^*$-theory description.  From the type IIB point of view, there is
no singularity in the metric as $H$ passes through zero (see
(\ref{hmet})), although $x^+$ becomes a null coordinate at this
transition point, which is why the T-duality transformed solution
becomes singular there.  On the other hand the type IIB pp-wave {\it
is} singular at $r=0$ (if $Q$ is non-zero), whilst this is perfectly
regular in the M-theory picture where it corresponds to the
AdS$_4\times S^7$ near-horizon limit.

    In the $r\longrightarrow 0$ limit the metric becomes AdS$_4\times
S^7$ and hence the supersymmetry is fully restored.  At large $r$, the
solution can be T-dualised to the pp-wave in type IIB, which preserves
at least half the supersymmetry.  Thus the deformed M2-brane is dual
to a three-dimensional field theory whose renormalisation group flow
runs from the conformal fixed point in the infra-red regime (at small
$r$) to a non-conformal theory as the energy increases.  At a certain
intermediate energy scale there is a phase transition associated with
the naked singularity of the M2-brane.  In the ultra-violet limit the
theory is related by T-duality to an exactly-solvable massive IIB string
theory.

\section{New supersymmetric deformed M2-branes}\label{newm2sec}

    We observed in section \ref{gensusysec} that whenever a pp-wave
in the type IIB theory has \nonstan Killing spinors, coming
from (\ref{susyfrac2}), these will be independent of $x^+$ and thus they
will survive in a T-dualisation and lifting to $D=11$ supergravity.
However, a naive application of the supersymmetry criterion (\ref{susycon})
to check whether these Killing spinors are present in the deformed M2-brane
solution will appear to lead to a contradiction, since (\ref{susycon})
will be violated.  In this section we shall discuss the resolution of
this puzzle, by showing how there are indeed Killing spinors in the
deformed M2-brane solution corresponding to the \nonstan Killing 
spinors in the type IIB pp-wave.  They are unusual in that they 
satisfy the $D=11$ Killing spinor equations by virtue of a cancellation 
between contributions from the extra flux $\Phi_\4$ and contributions from
the spin-connection and standard M2-brane-charge terms in the supercovariant
derivative.  Since they therefore cannot be viewed as Killing spinors 
that existed already in the undeformed M2-brane, whose ``survival''
under the deformation is then being tested, the assumptions that were made in 
the derivation of the standard supersymmetry criterion (\ref{susycon}) are
not valid.

   We shall illustrate this point by discussing in detail the example
of the Penrose limit of AdS$_5\times S^5$, which gives the pp-wave
with the maximal number 16 of \nonstan Killing spinors
\cite{blafighulpap}.  The calculations for the other pp-waves that we
have been considering in this paper are very similar.

   Substituting the pp-wave background (\ref{deformedm2}) with a
general self-dual 4-form $\Phi_\4$ into the
gravitino transformation rule $\delta\psi_M={\cal
D}_M\, \ep$, with
\be
{\cal D}_M = \nabla_M - \fft1{288}\, \Big( F_{N_1\cdots N_4}\, 
\Gamma_M{}^{N_1\cdots N_4} - 8 F_{M N_1\cdots N_3}\, 
\Gamma^{N_1\cdots N_3}\Big)\,,
\ee
we find that
\bea
{\cal D}_\m &=& \del_\mu -\ft16 H^{-3/2}\, \del_i H\, \Gamma_\mu\, 
\Gamma_i\, (1-\Gamma) - \ft1{12}\mu\,  
H^{-1}\, \Gamma_\mu\, W\,,\label{mudir}\\
{\cal D}_i &=& \del_i + \ft1{12} H^{-1}\, \del_j H\, \Gamma_{ij}\, 
(1-\Gamma) + \ft16 H^{-1}\, \del_i H\, \Gamma +\ft1{24}\mu\, H^{-1/2}\, 
(\Gamma_i\, W - 3 W\, \Gamma_i)\,,\label{idir}
\eea
where $W$ is given by (\ref{wdef}) and $\Gamma\equiv 
\ft16 \ep_{\mu\nu\rho}\, \Gamma^{\mu\nu\rho}$.  (Note that all indices
on $\Gamma$ matrices are tangent-frame indices.) 

    For the specific case of the deformed M2-brane corresponding
to the T-dualisation of the Penrose limit of AdS$_5\times S^5$
the self-dual 4-form $\Phi_\4$ given by (\ref{2term}), and 
$W$ for this example, given by (\ref{2termw}), 
has the following properties:
\be
W^2=2(1+\Gamma)\,,\quad W\, \Gamma=\Gamma\, W=W\,.
\ee
One can now verify, using these, that the following spinors 
$\ep$ satisfy the Killing-spinor condition ${\cal D}_M\, \ep=0$:
\be
\ep = H^{-1/6}\, (\del_i H\, W\, \Gamma_i - H^{1/2})\, \eta
\ee
where $\eta$ is a constant spinor satisfying $(1+\Gamma)\, \eta=0$ 
(and hence $W\, \eta=0$).  In addition, the function $H$ must satisfy
\be
\del_i\, \del_j\, H=-\ft18 \mu^2\, \delta_{ij}\,,
\ee
or, in other words,
\be
H= c_0 - \ft1{16} \mu^2\, z_i^2\,.
\ee
Thus we have verified that there are 16 Killing spinors for this 
deformed M2-brane solution, which is precisely the one related by 
T-duality to the pp-wave that is the Penrose limit of AdS$_5\times S^5$.

    It should be emphasised that the supersymmetry in this solution is
achieved by ``trading off'' the contributions from the extra 4-form flux 
(the $W$ terms in (\ref{mudir}) and (\ref{idir})) against the 
$\del_i H$ terms that usually cancel by themselves in the supersymmetry
transformation rules for a deformed M2-brane.  It is for this reason 
that the usual supersymmetry criterion (\ref{susycon}) is inapplicable in
this special case.

  It is straightforward to repeat the above analysis for all the other
examples of pp-waves with \nonstan Killing spinors that we have
constructed in this paper.  We shall not present the details here, since
the manipulations are very similar.  Furthermore, since the 
$x^+$-independence of the \nonstan Killing spinors implies that
they must survive as Killing spinors in the T-dualised picture, our
illustrative example above suffices to establish the principle of how
this can happen, despite the fact that the standard supersymmetry
criterion (\ref{susycon}) is not satisfied.

   We can also use the results in this section to study the  
supersymmetries of the deformed M2-branes related by T-duality to the  various 
pp-waves obtained in section (\ref{xpsec}) that have $x^+$-independent 
\stan Killing spinors, with the 4-form $\Phi_\4$ 
given by (\ref{ledf}).  In these cases there is no longer any 
``trading off'' between the $W$ terms and the $\del_i H$ terms 
in (\ref{mudir}) and (\ref{idir}), and hence any Killing spinors must be 
annihilated by these two types of term separately.  This means that
the supersymmetry criterion (\ref{susycon}) applies in these examples, and
so the Killing spinors are simply given by $\ep= H^{-1/6}\, \eta$, where
$\eta$ is any constant spinor satisfying $(1-\Gamma)\, \eta=0$ and 
(\ref{susycon}) (which can be expressed as $[\Gamma_i,W]\, \eta=0$).
One can now verify that there are $2n$ such Killing spinors if any $n$ 
of the equations (\ref{meqs}) are satisfied, with $1\le n\le 6$.  This is 
exactly in accordance with our supersymmetry discussion  for the pp-waves
in section \ref{xpsec}, where we counted the subset of the 16 \stan
Killing spinors that were independent of $x^+$.

   It is interesting to compare our results for these supersymmetric
deformed M2-branes with the results in \cite{duevkhlumi}, where a
self-dual 4-form with a structure contained within (\ref{ledf}) was
considered.  In \cite{duevkhlumi} the 4-form $L_\4$ was taken to be
the $G_2$-invariant structure constants in the multiplication table of
the imaginary octonion units, and the solution was found to be
non-supersymmetric.  This corresponds to all the seven constants $m_\a$
having unit magnitude, $|m_\a|=1$, and it is then evident from
(\ref{meqs}) that none of the eigenvalues $\lambda_i$ will vanish,
and thus none of the 16 \stan Killing spinors will be independent
of $x^+$.  However, we have also seen that for any of the pp-waves
constructed using (\ref{ledf}), there will be at least two \nonstan
Killing spinors, provided that one takes $Q=0$ in (\ref{isotropic}), 
and chooses the constants $\mu_i$ to satisfy (\ref{muicon}).  For the
particular example for $\Phi_\4$ considered in \cite{duevkhlumi}, 
we find that the $\lambda_i$ in (\ref{meqs}) are given by
\be
\lambda_i= (14,-2,-2,-2,-2,-2,-2)\,,
\ee
and so the smallest degeneracy is $k=1$, implying exactly two 
\nonstan Killing spinors.  It follows, therefore, that with the
choice for $\Phi_\4$ made in \cite{duevkhlumi}, the ``octonionic M2-brane''
will have two supersymmetries if $H$ is taken to have the non-isotropic
form 
\be
H = c_0 -\ft{1}{16} \mu^2\, (49 z_1^2 + z_2^2 + z_3^2 +\cdots + z_8^2)\,.
\ee

   It is also worth remarking that we can obtain further examples of
supersymmetric deformed M2-branes by taking $\Phi_\4$ to be harmonic
and given by $\Phi_\4= r^{-8}\, (L_\4 + {*L_\4})$, with $L_\4$ again
given by (\ref{ledf}).  The fractions of preserved supersymmetry are
again governed by (\ref{meqs}).  For the solutions with
$\Phi_\4= r^{-8}\, (L_\4 + {*L_\4})$ the 4-form is square integrable
at large distance but divergent at small distance. The resulting
solution is accordingly well-behaved at large distance, but has a
naked singularity at small distance.  The singularity can be resolved
by replacing the flat transverse space by a Ricci-flat space with
special holonomy that admits an $L^2$-normalisable harmonic 4-form
\cite{clptrans,cglpsten,cglphyper,newspin7}.  By dimensional reduction
and T-duality we can then obtain pp-waves in type IIB supergravity
where the flat 8-metric is replaced by the space of special holonomy.
However, although these pp-waves are non-singular, there is no reason
to expect that these backgrounds would correspond to exactly solvable
string theories. 

\section{Conclusions}

   In the paper we have obtained a large class of pp-waves in type IIB
supergravity theories, in which a constant five-form characterised by
seven parameters is turned on.  These solutions in general lead to
exactly solvable string backgrounds.  Our principal focus was on the
analysis of the number of surviving supersymmetries. In addition to 16
``\stan supersymmetries,'' these backgrounds always allow for the
possibility of further ``\nonstan supersymmetries.''  The conditions
for the appearance of \nonstan supersymmetries precisely determines
that the masses of the world-sheet bosonic fields must match those of
the fermionic fields, thus ensuring linearly-realised world-sheet
supersymmetry of the corresponding string action.

    The analysis of the above class of pp-wave solutions demonstrates
a one-to-one correspondence between the \nonstan supersymmetry and the
world-sheet supersymmetry.  The Penrose limit of $AdS_5\times S^5$
is a particular example in this class, with all eight world-sheet
superfields having equal and non-vanishing masses.  We also found
another example of a Penrose limit, namely the Penrose limit of
$AdS_3\times S^3$ space, which corresponds to the near horizon limit of
a D3/D3 intersection. For this example, the \nonstan supersymmetry determines
that four world-sheet superfields have zero masses, and the other four
have equal and non-zero masses.

    The pp-waves can all be T-dualised and then lifted to
eleven-dimensions, where they become ``deformed M2-branes'' with an
additional 4-form flux.  Any Killing spinor in the pp-wave that is
independent of the $x^+$ T-dualising coordinate will necessarily
continue to be a Killing spinor in the $D=11$ supergravity picture.
The 16 \stan Killing spinors in the pp-wave are always independent of
$z^i$, but in general depend on $x^+$ unless they are annihilated by
$W$.  By contrast, any \nonstan Killing are automatically independent
of $x^+$, but they depend in a non-trivial way on the $z^i$
coordinates.  After the T-dualisation, any of the 16 \stan Killing spinors
that are $x^+$-independent give rise to Killing spinors of the usual sort
in the deformed M2-brane, which satisfy the standard supersymmetry criterion
(\ref{susycon}).   By contrast, all the \nonstan Killing spinors give rise
to Killing spinors in the deformed M2-brane for which the standard 
supersymmetry criterion (\ref{susycon}) is not satisfied.  They are 
annihilated by the $D=11$ supercovariant derivative because of a cancellation
between terms involving the extra 4-form flux and terms from the
spin connection and the usual 4-form contribution.

   We also observed that by starting from any deformed M2-brane
solution in $D=11$, with $dz_i^2$ replaced by an 8-metric $ds_8^2$ of
special holonomy and $\Phi_\4$ a self-dual harmonic 4-form in
$ds_8^2$, we can obtain a supersymmetric pp-wave solution in the type
IIB theory, by means of a dimensional reduction to type IIA and then a
T-duality transformation.  The resulting pp-wave will have 16
``standard'' Killing spinors, regardless of whether or not the
deformed M2-brane is supersymmetric.  There will not be any
supernumerary Killing spinors in these generalised pp-waves.  If the
deformed M2-brane has supersymmetries, then the corresponding subset
of the 16 standard Killing spinors in the T-dual picture will be
independent of $x^+$. The deformed M2-brane solutions could be of the
type principally considered in
\cite{hawtay,clptrans,kbec,cglpsten,cglphyper,newspin7}, where the
self-dual harmonic 4-form $\Phi_\4$ is square-integrable, or else one
could take $\Phi_\4$ to be a non-normalisable self-dual harmonic
4-form, such as the covariantly-constant associative 4-form in a space
of Spin(7) holonomy, or $J\wedge J$ in a Ricci-flat K\"ahler 8-metric.

\section*{Acknowledgements}

   We are grateful to Gary Gibbons and Jim Liu for conversations.
H.L.~and C.N.P.~are grateful to UPenn for hospitality and financial
support during the course of this work.

\end{document}